\documentclass[10pt,aps,pra,showpacs,twocolumn]{revtex4-1}
\usepackage{amsmath,amssymb,graphicx}
\usepackage{mathptmx}
\usepackage{color}

\begin{document}

\title{Spatiotemporal light beam compression from nonlinear mode coupling}


\author{Katarzyna Krupa$^1$}
\email{katarzyna.krupa@unibs.it}
\author{Alessandro Tonello$^2$}
\author{Vincent Couderc$^2$} 
\author{Alain Barth\'el\'emy$^2$}
\author{Guy Millot$^3$}
\author{Daniele Modotto$^1$}
\author{Stefan Wabnitz$^{1,4}$}

\affiliation{$^1$Dipartimento di Ingegneria dell'Informazione, Universit\`a di Brescia, via Branze 38, 25123, Brescia, Italy}
\affiliation{$^2$Universit\'e de Limoges, XLIM, UMR CNRS 7252, 123 Av. A. Thomas, 87060 Limoges, France}
\affiliation{$^3$Universit\'e de Bourgogne, ICB, UMR CNRS 6303, 9 Av. A. Savary, 21078 Dijon, France}
\affiliation{$^4$ Istituto Nazionale di Ottica del Consiglio Nazionale delle Ricerche (INO-CNR), Via Branze 45, 25123 Brescia, Italy}

\date{\today}

\begin{abstract}
We experimentally demonstrate simultaneous spatial and temporal compression in the propagation of light pulses in multimode nonlinear optical fibers. We reveal that the spatial beam self-cleaning 
recently discovered in graded-index multimode fibers is accompanied by significant temporal 
reshaping and up to four-fold shortening of the injected sub-nanosecond laser pulses.  
Since the nonlinear coupling among the modes strongly depends  
on the instantaneous power, we explore the entire range of the nonlinear dynamics with a single optical pulse, 
where the optical power is continuously varied across the pulse profile.

\end{abstract}


\maketitle


\section{Introduction}
Methods to shape and control the propagation of electromagnetic radiation are of great importance in various fields of science and technology, such as atomic and plasma physics, communications, material processing, and biomedicine \cite{wein09,forbe14}. Among these, of particular interest are methods that enable the simultaneous control of spatial and temporal degrees of freedom of wave packets. 
To this end, two different strategies have been exploited. The first methodology is based on a spatiotemporal synthesis of a special input wave, so that diffractive and dispersive effects compensate for each other upon {\em linear} propagation in the material. 
Building blocks of these linear light bullets are Bessel beams and their linear combinations, along with Airy pulses, leading to spatiotemporal invariant packets 
\cite{bibDurnin87} 
The second approach involves the generation of solitary waves, that exploit the {\em nonlinear} (quadratic or cubic) response of the material for compensating diffractive and dispersive wave spreading. Although successfully exploited in (1+1)D propagation models, 
in more than one dimension spatiotemporal solitons have so far largely eluded experimental observation, owing to the presence of 
modulation instability (MI), collapse and filamentation \cite{wise02}. 


In the least few years, an alternative approach has emerged to control the spatiotemporal properties of a light beam, based on complex nonlinear mode mixing in multimode optical fibers (MMFs) \cite{Wright2015R31}. 
MMFs permit, by managing the number of guided modes, to vary the dimensionality of wave propagation from the limit case of (1+1)D (single mode fibers), up to the 
free-space or (3+1)D case (highly multimode fibers) \cite{Wright2015R29}. 
In spite of the maturity of the field of nonlinear fiber optics, new intriguing wave propagation phenomena
\cite{Mafi2012R36}, such as multimode solitons \cite{Renninger2012R31} and ultra-wideband sideband series \cite{Wright2015R31, Wright2015R29,KrupaPRLGPI}, have been observed only recently in MMFs. 
It is well known that linear wave propagation in MMFs  leads to spatial spreading of a light beam among 
a multitude of guided modes, which results in highly irregular speckled patterns at the fiber output, an effect known as modal noise in the context of fiber optics communications.

Work by Krupa et al. \cite{Krupanatphotonics} led to the unexpected discovery that the intensity dependent contribution to the refractive index, or Kerr effect, in a graded index (GRIN) MMF has the capacity to counteract modal noise, and lead to the formation of a highly stable, spatially compressed beam close to the fundamental fiber mode. This spatial beam cleaning is typically observed over a few meters of GRIN MMF at threshold power levels (of the order of 1 kW) that are three orders of magnitude lower than the value for catastrophic self-focusing, and in a quasi-continuous wave (CW) propagation regime (i.e., by using sub-nanosecond pulses), so that dispersive effects can be neglected \cite{Krupanatphotonics,KrupaPRLGPI,WrightNP2016,Galmiche:16,KrupaLuot:16}. Spatial beam cleanup using femtosecond pulses 
has also been reported \cite{LiuKerr}, however at power levels which are about two orders of magnitude larger than in the quasi-CW regime, so that dispersive and self-focusing effects may play a significant role in this case. Whereas in the quasi-CW regime, nonlinear mode coupling is the sole mechanism responsible for the cleaning of the 2D transverse spatial beam profile.
In this situation, the temporal pulse self-reshaping that accompanies 
spatial beam cleaning has not yet, to the best of our knowledge, been investigated. Hence it remains an open problem. 

In this paper, we fil this gap, by experimentally revealing the complex phenomenology associated with the temporal dimension of tKerr beam self-cleaning.
A particularly important consequence of temporal reshaping is that, for input pulse power values slightly larger than the 
threshold value for spatial beam self-cleaning, the sub-nanosecond laser 
pulses may undergo significant temporal compression, {\it i.e.}, a time shortening, accompanied by peak power enhancement. Our theoretical analysis of nonlinear mode coupling confirms well the observed complex spatiotemporal reshaping. 

These findings are of great importance for the design of high power fiber amplifiers \cite{GuenardOpex} and mode-locked lasers \cite{GuenardOpex2,WiseScience} based on MMFs, and their technological applications. The experimentally observed spatial beam reshaping into a quasi-fundamental mode of the nonlinear multimode fiber, accompanied by temporal pulse compression, is analogous to the discrete spatiotemporal focusing and compression, which occurs with pulses propagating in multicore optical fibers 
\cite{PhysRevLett.105.263901,Rubenchik:15,PhysRevA.94.043848}. 

From a fundamental perspective, our results have a broad interest, by providing an important example of spatio-temporal self-organization in a complex system: laws governing the collective evolution of the guided modes are not predictable in terms of the evolution of individual modes.
Moreover, our results have implications across different fields such as quantum fluids of light \cite{Caru2013} and matter waves in atomic systems, where Bose-Einstein condensation in the presence of a trapping potential is described in terms the Gross-Pitakevskii equation, formally equivalent to the (3+1)D nonlinear Schr\"odinger equation ruling propagation in GRIN MMFs \cite{PRA11b}.

\section{Experimental Results}

\begin{figure}
\includegraphics[width=\linewidth]{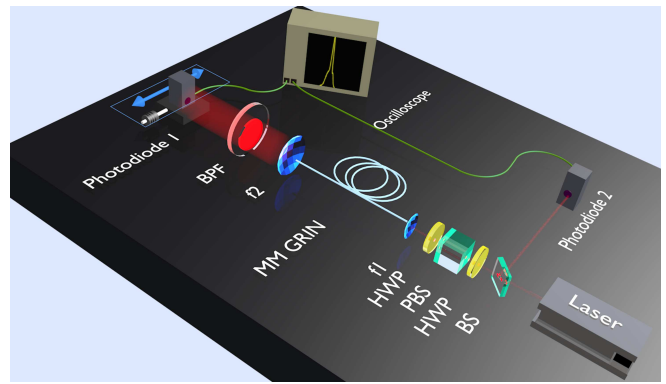}
\caption{Experimental setup: the laser beam is divided by a  beam splitter (BS) to provide a trigger reference time.
The beam directed to the fiber passes first through two half wave plates (HWP) and a polarization beam splitter 
cube (PBS). Input and output lenses have focal lengths f1=50~mm and f2=8~mm. The output beam is filtered by a band-pass filter (BPF) of 10~nm bandwidth centered at 1064~nm. \label{fig:setup}}
\end{figure}

In our experimental setup (see fig. \ref{fig:setup}), we pumped the fiber with an 
amplified microchip Q-switched laser delivering pulses of 740~ps at a repetition rate of 27~kHz, with a central wavelength of 1064~nm. The maximum energy per pulse reaches more than 100~$\mu J$, 
with a an instantaneous 
peak power level up to 150~kW. 
A fast photodiode (In-GaAs PIN Detector ET-3500, with the rise time of 25 ps, photodiode 1 in fig.1) was connected to a 20-GHz bandwidth oscilloscope and used to measure the output pulse profile with a trigger signal coming from photodiode 2.
In fact, due to the pulse-to-pulse fluctuations in the laser emission time for this type of source, it was necessary to resort to self-referenced temporal measurements. Namely, a small fraction of the emitted laser beam was used as trigger signal, and sent through a beam splitter (BS) to a second fast photodiode (photodiode 2). 
The major fraction of the beam power was transmitted by the BS to a 
Glan prism (or polarisation beam splitter PBS), 
placed in between two half-wave plates (HWPs) at 1064~nm to control the maximum input peak power. 
This configuration permitted us to control the maximum input peak power, as well as the orientation of linear state of polarisation of
laser pulses. 
The optical beam was coupled into a 12-m long GRIN 50/125 multimode fiber
by means of a positive lens (f1), that delivers a Gaussian beam with a diameter of 40~$\mu m$ at the input end of the fiber. 
Light at the fiber output was imaged into the detection system by means of lens f2, after spectral selection by a 3-nm interferometric band pass filter (BPF) with the central wavelength of 1064~nm, in order to select the pump beam only. 

Photodiode 1 had a much smaller effective area than the dimension of the imaged field (which was magnified x60). 
The diameter of the active window of the photodiode will appear 
of only  4~$\mu m$ if one 
rescales the magnified image of the near field into the 50~$\mu m$
diameter  of the fiber core. 
Temporal waveforms could be detected in real time in our setup, with samples spaced by 25~ps, 
which is roughly 30 times shorter than the input pulse duration.
\begin{figure}
\includegraphics[width=\linewidth]{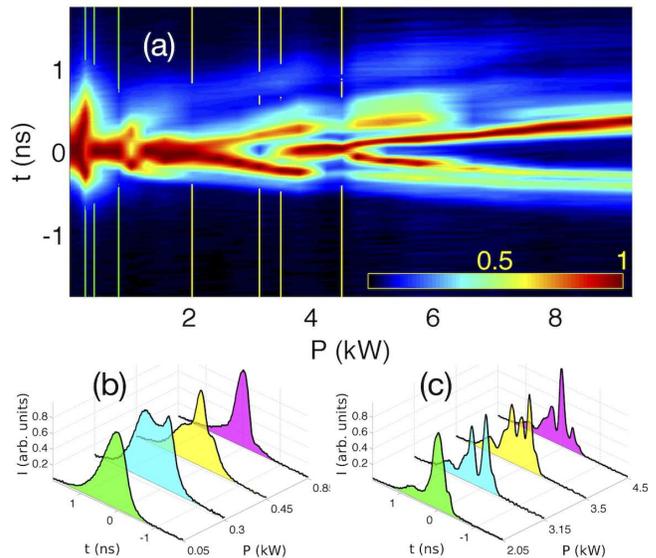}
\caption{Experimental results: (a) output temporal waveforms upon input peak power. 
Vertical lines identify some peak power levels, whose corresponding waveforms are reported 
for ease of comparison in panels (b) and (c).  \label{fig:oscilloscope}}
\end{figure}

In fig. \ref{fig:oscilloscope} we show a collection of output temporal waveforms that were obtained for different input peak powers.  
In these measurements, the active area of photodiode 1 was 
placed at the center of the near field image of the beam emerging from the fiber output. In this configuration,
the photodiode mainly detects light carried by fiber modes with a maximum in their center (e.g., the fundamental mode). Figure \ref{fig:oscilloscope}(b) shows that, for input peak power values below 50~W, the pulse envelope 
reproduces the pulse waveform of the microchip laser. A first temporal pulse distortion appears as the input peak power is increased up to 0.3~kW: as can be seen, the top of the pulse has a dip in the center. 
Note that the influence of chromatic and modal dispersion effects on pulse propagation is negligible, owing to the long pulse duration and the short length of the fiber. 
By further rising the input power up to 0.8~kW, a single compressed peak (sitting on low power wings) forms at the center of the output pulses. The full width at half maximum pulse duration is nearly halved (from 740~ps down to 425~ps) with respect to the input pulse. Upon subsequent increases of the input power, the temporal reshaping of the pulse envelope exhibits a recurring behavior, composed by a series of temporal broadenings, followed by the formation of
a dip at the pulse center and then the growth of a narrower, high power peak.
Figure \ref{fig:oscilloscope}(c) shows that, for an input peak power of 4.5~kW, the output pulse duration compresses down to only 175~ps, which is more than 4-fold shorter 
than the input pulse duration. 
Such pulse duration
is 7 times longer than the temporal spacing between samples of our measurement system. 
Correspondingly, the output peak power is about twice higher than the input value, since half of the energy is estimated to remain in the low power pulse wings.
For input peak powers of 5~kW and above, efficient Raman conversion (roughly up to 30\%) occurs from the input pump pulse into a Stokes wave. 
This is also enhanced by the progressive growth of peak power along the fiber. 
Raman scattering leads to a power-dependent energy depletion of the output pulses
measured at 1064~nm.

\begin{figure}
\includegraphics[width=\linewidth]{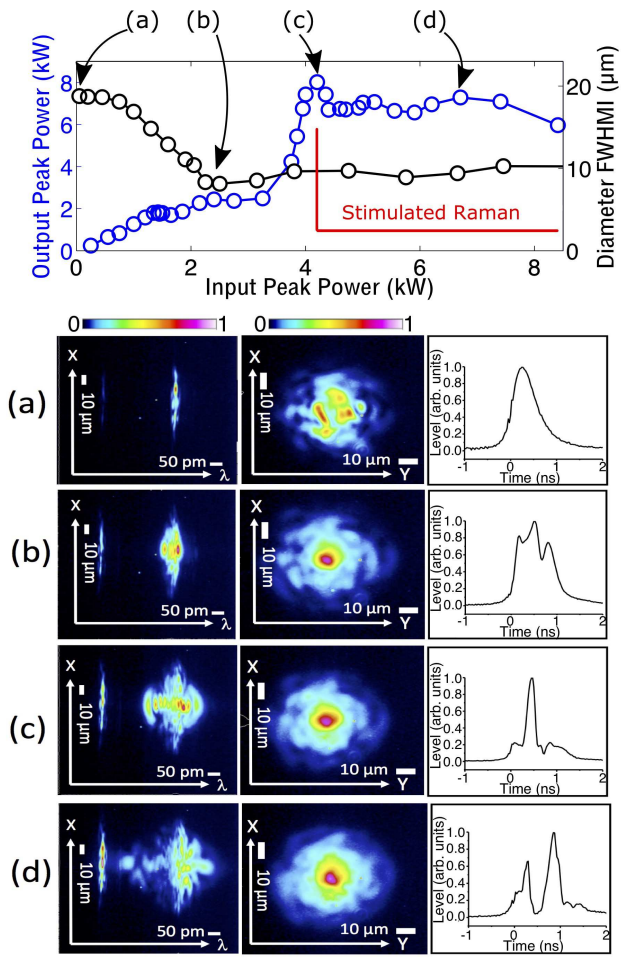}
\caption{Experimental results. The curves in the top panel show the output peak power
and the corresponding beam diameter upon input peak power. The markers (a)-(d) represent four characteristic points for which the lower panels report the spectrum, the spatial beam shape and the corresponding temporal profile measured by photodiode 1. \label{fig:synthesis}}
\end{figure}

To fully characterize the nonlinear spatial, temporal and spectral beam reshaping at the MMF output, we carried out a comparative analysis as shown in fig. \ref{fig:synthesis}. Here we report, by the use of different measurement devices, the power dependence of the output beam across its different dimensions. 

The two curves on the top panel of fig. \ref{fig:synthesis} illustrate the evolution of the output beam diameter and pulse peak power, as a function of input peak power. 
The output pulse peak power was calculated by combining the pulse temporal shape from photodiode 1 with the measurement of the average output power from a calibrated power meter. 
The temporal signal is proportional to the beam intensity spatially integrated over the active surface of the photodiode.
Arrows with labels (a-d) in the top panel of fig. \ref{fig:synthesis} refer to four characteristic values of the input peak power, with corresponding spatial, spectral and temporal measurements appearing in panels (a)-(d) of fig. \ref{fig:synthesis}. 
Measurements in the left column of these panels show the diameter of the output beam (across a fixed vertical spatial dimension) as a function of wavelength, and were taken by a dispersive spectroscope \cite{bibspectroscope}, a home-built instrument 
permitting us to obtain an image of the spectrum with a resolution lower than 50~pm. The center column of these panels shows the near field of the output spatial beam, measured with a CCD camera. Finally, the right column of panels (a)-(d)
provides the corresponding temporal envelope of the output pulses. 

Panels of group (a) in fig. \ref{fig:synthesis} illustrate the spatiotemporal properties of the output beam at the lowest measured input peak power. The spectrum clearly displays the presence of the main longitudinal mode of the microchip laser. The camera reveals a speckled beam, whereas the oscilloscope displays an output pulse shape which is very close to the input pulse from the laser.
As demonstrated by fig. \ref{fig:oscilloscope}(a-b), nonlinear temporal reshaping of the output pulse occurs for input power levels above 0.3~kW, which is well below the threshold of spatial self cleaning
(of about 1~kW in a 12-m long fiber \cite{Krupanatphotonics}). 
This indicates that significant nonlinear mode coupling already occurs in the fiber at power levels that are below the observation of output spatial beam reshaping.
For an input power level of 2~kW, panels (b) in fig. \ref{fig:synthesis} show that the transverse profile of the output beam exhibits self-cleaning, namely, a bell-shaped beam close to the fundamental mode of the GRIN fiber is formed (see Ref.\cite{Krupanatphotonics} for details).
The temporal waveform in the corresponding panel of fig. \ref{fig:synthesis}(b) unveils the presence of a strong temporal modulation. Whereas the dispersive spectroscope
reveals that only slight spectral broadening is induced by self-phase modulation. 
Incidentally, the signal from the dispersive spectroscope displays an additional spectral peak, 
which is only due to the second longitudinal mode of the microchip laser.  

As can be seen from fig. \ref{fig:synthesis}(b-d), as the input power is increased from 2 up to 7~kW, the spatial profile of the output beam maintains a stable self-cleaned transverse profile. However we  observe a progressive spectral broadening, which is associated with output pulse narrowing from 740~ps down to 175~ps (panels of group (c)). Note that, with an input peak power of 4~kW, we estimate (by combining oscilloscope and average output power measurements) that the output peak power doubles to 8~kW.  
Panels in fig. \ref{fig:synthesis}(d) reflect the additional contribution of stimulated Raman scattering (SRS): differently from previous cases at lower power levels, where spectral broadening was within the bandwidth of BPF, in this case 
 energy depletion in the pulse temporal center is observed in presence of Raman conversion. 
 Despite the transfer into the Stokes wave
 of a part of the beam energy, 
 spatial beam cleaning remains active. 

\begin{figure}
\includegraphics[width=\linewidth]{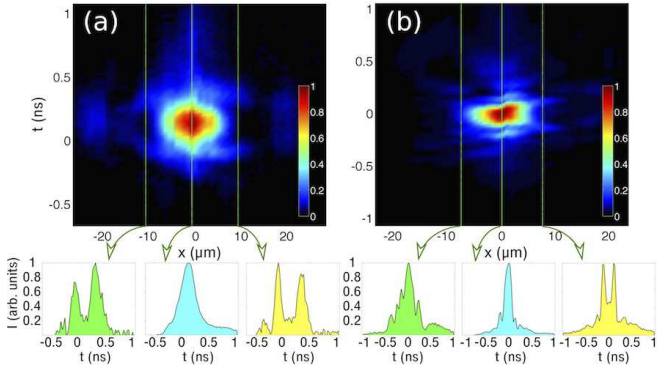}
\caption{Experimental temporal waveforms for different transverse positions (x). The input peak powers were
2~kW for panel (a), and 4~kW for panel (b). In both cases the color bar has been referred to the maximum signal level measured at the center of the beam. The vertical lines are guides to the eye to identify specific measurement points whose associated temporal waveforms are shown by the corresponding insets, with intensity normalized to its local maximum level.    \label{fig:3B}}
\end{figure}

Since the active area of the photodiode has a diameter 12 times smaller than the diameter of the near field image, it is of fundamental importance to explore the spatiotemporal properties of the output beam at different positions of the beam transverse profile. For this aim, we shifted along the horizontal coordinate (x axis) the relative position of the photodiode with respect to the beam image, while keeping the vertical position (y) fixed. In this way we may explore the intermediate and the maximum stages of spatial beam cleaning, before SRS conversion takes place.  
Figure \ref{fig:3B} summarizes these results by showing the temporal profiles of the output field that are obtained when moving the photodiode along the horizontal (x) spatial coordinate. Here panels (a) and (b) are obtained for input peak powers of 2~kW  or 4~kW, respectively. The vertical lines  highlight illustrative transverse spatial positions x: the associated temporal waveforms, normalized to their local maximum levels, are shown by the corresponding insets.  These observations 
confirm that light intensity along the temporal pulse envelope
can be localized in the transverse domain in different ways, so that space 
and time cannot be separated.    
  
We also used long series of similar measurements (here not shown) in order to characterize the spatial
distribution of light. Namely, we checked the agreement between the integrated intensity of time profiles
measured at different output positions x, with the spatial beam profile that is obtained from a CCD camera, 
which integrates the signal power in the time domain, owing to its millisecond (or tens of millisecond) response time.
Spatial beam cleaning is apparent in both spatiotemporal diagrams 
and in two-dimensional images of the output transverse beam profile. 
However, for a 4~kW input power level 
spatial cleaning extends further across the low power background surrounding the central bell-shaped lobe, and 
a net compression 
of the temporal envelopes is observed 
at larger distances from the beam center.

\begin{figure}
\includegraphics[width=\linewidth]{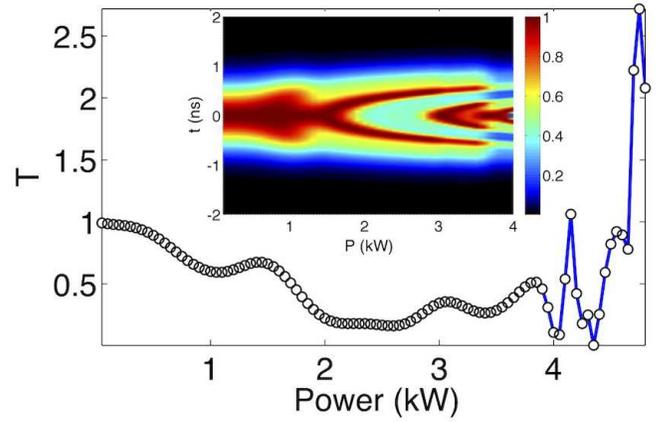}
\caption{Numerical results from eq. \ref{eq:modexp} with 16 modes, where only the first 9 were excited at the input and the remaining modes set to  zero. 
The curve represents the coefficient $T=|a_0(L)|^2/|a_0(0)|^2$ for different input peak power levels. This curve can be considered as a transfer function representing the output contribution of the fundamental mode at different input power values. 
The inset shows the evolution of the temporal waveform associated to the fundamental mode $a_0$
when assuming an input pulse duration of 1~ns.
 \label{fig:numerics}}
\end{figure}

\section{Numerical Results}

We studied spatiotemporal beam reshaping in the GRIN MMF in terms of nonlinear mode coupling equations for the guided mode amplitudes $b_n(z)=a_n(z)exp(i\beta_n z)$
\begin{equation}
\frac{da_n}{dz}=i\gamma \sum_{m,p,q}Q_{m,p,q,n}a_ma_p^*a_q\exp{(i\Delta\beta_{m,p,q,n} z)}
\label{eq:modexp}
\end{equation}

The function $a_n$ represents the slowly varying complex amplitude of mode $n$, with corresponding linear propagation constant $\beta_n$. $\Delta\beta_{m,p,q,n}=\beta_m-\beta_p+\beta_q-\beta_n$ is the phase mismatch for the mode group with indices $m,n,p,q$ and overlap coefficients $Q_{m,p,q,n}=\iint\psi_m\psi^*_p\psi_q\psi_n^*dxdy$, being $\psi_n(x,y)$ the n-th element 
of the orthonormal basis of guided modes. 
$\gamma=n_2\omega/c$ is the nonlinear coefficient of the fiber at the angular frequency $\omega$, with $n_2$ the nonlinear Kerr parameter of silica. 
Note the absence of time 
derivatives in eq.\ref{eq:modexp}: since dispersion is not relevant for the considered fiber length and the pulse durations, the time dependence 
observed in the experiments can be modeled by simply solving  eq. \ref{eq:modexp} 
for different input power levels.
Spatial modes in GRIN fibers have the property of having nearly equally spaced propagation constants. Therefore, the phase mismatch values can also be arranged 
in a discrete set of values. 

To reduce computational burden, we limited our analysis to the first 16 modes of the Hermite-Gauss basis, and we integrated eq.\ref{eq:modexp} with a standard Runge-Kutta method, by only considering terms corresponding to coupled modes with near zero relative mismatch. Namely, for $|\Delta\beta_{m,p,q,n}|>S$, the corresponding overlap coefficient $Q_{m,p,q,n}$ was set to zero (S=50~rad/m in simulations).  The input mode power fraction and phase were first randomly chosen, and then kept constant as the input pulse power was varied.
Figure \ref{fig:numerics} illustrates the ratio T of the output power of the fundamental mode, 
normalized to its input value, as a function of
the input pulse peak power, at the end of a  12~m GRIN fiber. 
Since dispersive effects can be neglected,
with a photodiode placed at the beam center, the curve in fig.\ref{fig:numerics}, 
may be used to reconstruct the output temporal waveform, under the approximation 
that the fundamental mode only is sampled.

The inset of fig.\ref{fig:numerics} shows the corresponding output temporal profile as a function of input power, for an input  gaussian pulse of 1~ns.
As we can see, if the peak power is in the range of 2-3~kW, the transmission T
of the fundamental mode is much 
smaller than for input powers below 0.5~kW, in qualitative agreement with the experimental results 
(when looking at the center of the beam) as shown by fig.\ref{fig:synthesis}.  
The numerical prediction of the cycles of pulse compression and breaking of fig.\ref{fig:numerics} are also in good agreement with the experimental result of fig.\ref{fig:oscilloscope}(a).

The explanation of the observed nonlinear transmission curves such as those of fig.\ref{fig:numerics}  can be given in the theoretical framework of the self-switching of pulses 
in nonlinear coupled mode devices. As previously discussed, the photodiode placed 
at the center of the output 
beam essentially samples the fraction of the fundamental mode emerging from the multimode fiber. 
Now, different portions of the long pulse self-switch independently of each other, and sample the same 
nonlinear CW transmission curve at different instantaneous power values. 
When the pulse peak power reaches a local maximum of the transmission curve, 
the output power 
in the center of the beam is also locally enhanced. 
The low-power 
pulse wings remain largely undistorted. This tendency is enhanced in experiments,
when the self-cleaning increases the fundamental mode content.   
This explains the reduction of the pulse temporal duration
that is observed 
in figs. \ref{fig:oscilloscope}, \ref{fig:synthesis}, and \ref{fig:numerics}.  
As the peak power grows larger, 
the wings of the pulse have now a sufficient power to experience self-cleaning into the fundamental mode. 
Whereas the power at the pulse center is so high, that the output spatial 
mode switches to another 
low order mode with a spatial distribution with dip at the center; 
the photodiode would detect this situation as a reduction
of instantaneous power. This explains the observation of temporal profiles with 
a hole in the center in figs.\ref{fig:oscilloscope}, \ref{fig:synthesis}, 
and \ref{fig:numerics}. 
The described phenomenology is universal to nonlinear coupled mode devices: for example, 
very similar nonlinear pulse reshaping occurs in the self-switching of a two-mode nonlinear birefringent fiber \cite{AssantoTrilloWabnitzAPL}.   

Furthermore, numerical simulations predict that, for input powers above 5~kW, 
the mode coupling process is characterized by 
an irregular behavior leading to 
sudden variations of the relative energy content in the fundamental mode transmission as the input power is slightly varied. Again, this 
situation is similar to the case of a two-mode nonlinear coupler \cite{AssantoTrilloWabnitzAPL}. In the experiments, nonlinear dissipation due to SRS-induced depletion of pump pulses (occurring for input peak powers above 4.5~kW) largely prevents the observation of this irregular pulse shaping. 

\section{Conclusions}

In conclusion, we have shown that complex mode mixing in nonlinear multimode fibers 
enables the simultaneous spatial cleanup in the transverse spatial dimensions, and  pulse compression in the temporal dimension. 
Spatial beam cleaning leads to the suppression of modal noise or irregular intensity fluctuations in the transverse dimension, and to the reshaping of the beam towards the fiber fundamental mode. 
The associated nonlinear mode coupling dynamics also results in significant temporal reshaping and pulse shortening, which may enhance the peak power of the output pulses. 
Our work should stimulate many other studies, including in particular a complete spatiotemporal beam characterization using more elaborate methods \cite{bib:trebinostripes}.

\begin{acknowledgments}
We acknowledge the financial support of: Horiba Medical and BPI france within the Dat@diag project; 
 iXcore research foundation; French National Research Agency ANR Labex ACTION; French program ''Investissement d'Avenir'', project ISITE-BFC-299 (ANR- 15-IDEX-0003); the European Research Council (ERC) under the European Union's Horizon 2020 research and innovation programme (grant agreement No.~740355).
K.~K. has received funding from the European Union's Horizon 2020 research and innovation programme under the Marie~Sk\l{}odowska-Curie grant agreement No.~713694 (MULTIPLY). 
\end{acknowledgments}

\end{document}